\documentclass[10pt, aps,prl,twocolumn,superscriptaddress, nofootinbib, longbibliography]{revtex4-1}[11pt]    	% revtex4-1 prc style

\pdfoutput=1
\usepackage{amsmath,amssymb,graphicx,lineno}
\usepackage{upgreek}
\usepackage{hyperref}
\usepackage{color}
\usepackage[utf8]{inputenc}
\usepackage{setspace}
\usepackage{lineno}
\usepackage{xcolor}
\usepackage{soul}
\usepackage{hyperref}

\usepackage{siunitx}
\DeclareSIUnit[number-unit-product = {}] \ev{eV}

\graphicspath{ {./plots/} }

\newcommand{\scix}[3][]
{
 \ifthenelse{\equal{#2}{}}
  {} % True i.e. No Mantissa
  { % False i.e. There is a Mantissa
   {#2} % Mantissa
  }
 \ifthenelse{\equal{#2}{}\OR\equal{#3}{}}
  {} % True i.e. No Times required
  { % False i.e. Times required
   {\hspace{-0em} \times \hspace{-0em}} % Times
  }
 \ifthenelse{\equal{#3}{}}
  {} % No Exponent
  {10^{#3}}   % Exponent
 \ifthenelse{\equal{#1}{}}
 {}
 {\,\mathrm{#1}} % Units, optional argument
}

\begin{document}
%\setstretch{2.5}

\title{
Nuclear Recoil Calibration at Sub-keV Energies in LUX \protect\\ and Its Impact on Dark Matter Search Sensitivity
}
\preprint{}

\author{D.S.~Akerib} \affiliation{SLAC National Accelerator Laboratory, 2575 Sand Hill Road, Menlo Park, CA 94205, USA} \affiliation{Kavli Institute for Particle Astrophysics and Cosmology, Stanford University, 452 Lomita Mall, Stanford, CA 94309, USA} 
\author{S.~Alsum} \affiliation{University of Wisconsin-Madison, Department of Physics, 1150 University Ave., Madison, WI 53706, USA}  
\author{H.M.~Ara\'{u}jo} \affiliation{Imperial College London, High Energy Physics, Blackett Laboratory, London SW7 2BZ, United Kingdom}  
\author{X.~Bai} \affiliation{South Dakota School of Mines and Technology, 501 East St Joseph St., Rapid City, SD 57701, USA}  
%\author{A.J.~Bailey} \affiliation{Imperial College London, High Energy Physics, Blackett Laboratory, London SW7 2BZ, United Kingdom}  
\author{J.~Balajthy} \affiliation{University of California Davis, Department of Physics, One Shields Ave., Davis, CA 95616, USA}  
\author{J.~Bang} \affiliation{Brown University, Department of Physics, 182 Hope St., Providence, RI 02912, USA}  
\author{A.~Baxter} \affiliation{University of Liverpool, Department of Physics, Liverpool L69 7ZE, UK}  
%\author{P.~Beltrame} \affiliation{SUPA, School of Physics and Astronomy, University of Edinburgh, Edinburgh EH9 3FD, United Kingdom}  
\author{E.P.~Bernard} \affiliation{University of California Berkeley, Department of Physics, Berkeley, CA 94720, USA}  
\author{A.~Bernstein} \affiliation{Lawrence Livermore National Laboratory, 7000 East Ave., Livermore, CA 94551, USA}  
\author{T.P.~Biesiadzinski} \affiliation{SLAC National Accelerator Laboratory, 2575 Sand Hill Road, Menlo Park, CA 94205, USA} \affiliation{Kavli Institute for Particle Astrophysics and Cosmology, Stanford University, 452 Lomita Mall, Stanford, CA 94309, USA} 
\author{E.M.~Boulton} \affiliation{University of California Berkeley, Department of Physics, Berkeley, CA 94720, USA} \affiliation{Lawrence Berkeley National Laboratory, 1 Cyclotron Rd., Berkeley, CA 94720, USA} \affiliation{Yale University, Department of Physics, 217 Prospect St., New Haven, CT 06511, USA}
\author{B.~Boxer} \affiliation{University of Liverpool, Department of Physics, Liverpool L69 7ZE, UK}  
\author{P.~Br\'as} \affiliation{LIP-Coimbra, Department of Physics, University of Coimbra, Rua Larga, 3004-516 Coimbra, Portugal}  
\author{S.~Burdin} \affiliation{University of Liverpool, Department of Physics, Liverpool L69 7ZE, UK}  
\author{D.~Byram} \affiliation{University of South Dakota, Department of Physics, 414E Clark St., Vermillion, SD 57069, USA} \affiliation{South Dakota Science and Technology Authority, Sanford Underground Research Facility, Lead, SD 57754, USA} 
%\author{S.B.~Cahn} \affiliation{Yale University, Department of Physics, 217 Prospect St., New Haven, CT 06511, USA}  
\author{M.C.~Carmona-Benitez} \affiliation{Pennsylvania State University, Department of Physics, 104 Davey Lab, University Park, PA  16802-6300, USA}  
\author{C.~Chan} \affiliation{Brown University, Department of Physics, 182 Hope St., Providence, RI 02912, USA}  
%\author{A.A.~Chiller} \affiliation{University of South Dakota, Department of Physics, 414E Clark St., Vermillion, SD 57069, USA}  
%\author{C.~Chiller} \affiliation{University of South Dakota, Department of Physics, 414E Clark St., Vermillion, SD 57069, USA}  
%\author{A.~Currie} \affiliation{Imperial College London, High Energy Physics, Blackett Laboratory, London SW7 2BZ, United Kingdom}  
\author{J.E.~Cutter} \affiliation{University of California Davis, Department of Physics, One Shields Ave., Davis, CA 95616, USA}  
%\author{T.J.R.~Davison} \affiliation{SUPA, School of Physics and Astronomy, University of Edinburgh, Edinburgh EH9 3FD, United Kingdom}  
\author{L.~de\,Viveiros}  \affiliation{Pennsylvania State University, Department of Physics, 104 Davey Lab, University Park, PA  16802-6300, USA}  
%\author{A.~Dobi} \affiliation{Lawrence Berkeley National Laboratory, 1 Cyclotron Rd., Berkeley, CA 94720, USA}  
%\author{J.E.Y.~Dobson} \affiliation{Department of Physics and Astronomy, University College London, Gower Street, London WC1E 6BT, United Kingdom}  
\author{E.~Druszkiewicz} \affiliation{University of Rochester, Department of Physics and Astronomy, Rochester, NY 14627, USA}  
%\author{B.N.~Edwards} \affiliation{Yale University, Department of Physics, 217 Prospect St., New Haven, CT 06511, USA}  
%\author{C.H.~Faham} \affiliation{Lawrence Berkeley National Laboratory, 1 Cyclotron Rd., Berkeley, CA 94720, USA}  
%\author{S.R.~Fallon} \affiliation{University at Albany, State University of New York, Department of Physics, 1400 Washington Ave., Albany, NY 12222, USA}  
\author{A.~Fan} \affiliation{SLAC National Accelerator Laboratory, 2575 Sand Hill Road, Menlo Park, CA 94205, USA} \affiliation{Kavli Institute for Particle Astrophysics and Cosmology, Stanford University, 452 Lomita Mall, Stanford, CA 94309, USA} 
\author{S.~Fiorucci} \affiliation{Lawrence Berkeley National Laboratory, 1 Cyclotron Rd., Berkeley, CA 94720, USA} \affiliation{Brown University, Department of Physics, 182 Hope St., Providence, RI 02912, USA} 
\author{R.J.~Gaitskell} \affiliation{Brown University, Department of Physics, 182 Hope St., Providence, RI 02912, USA}  
%\author{V.M.~Gehman} \affiliation{Lawrence Berkeley National Laboratory, 1 Cyclotron Rd., Berkeley, CA 94720, USA}  
%\author{J.~Genovesi} \affiliation{University at Albany, State University of New York, Department of Physics, 1400 Washington Ave., Albany, NY 12222, USA}  
\author{C.~Ghag} \affiliation{Department of Physics and Astronomy, University College London, Gower Street, London WC1E 6BT, United Kingdom}  
%\author{K.R.~Gibson} \affiliation{Case Western Reserve University, Department of Physics, 10900 Euclid Ave, Cleveland, OH 44106, USA}  
\author{M.G.D.~Gilchriese} \affiliation{Lawrence Berkeley National Laboratory, 1 Cyclotron Rd., Berkeley, CA 94720, USA}  
%\author{E.~Grace} \affiliation{Pennsylvania State University, Department of Physics, 104 Davey Lab, University Park, PA  16802-6300, USA}  
\author{C.~Gwilliam} \affiliation{University of Liverpool, Department of Physics, Liverpool L69 7ZE, UK}  
\author{C.R.~Hall} \affiliation{University of Maryland, Department of Physics, College Park, MD 20742, USA}  
%\author{M.~Hanhardt} \affiliation{South Dakota School of Mines and Technology, 501 East St Joseph St., Rapid City, SD 57701, USA} \affiliation{South Dakota Science and Technology Authority, Sanford Underground Research Facility, Lead, SD 57754, USA} 
\author{S.J.~Haselschwardt} \affiliation{University of California Santa Barbara, Department of Physics, Santa Barbara, CA 93106, USA}  
\author{S.A.~Hertel} \affiliation{University of Massachusetts, Amherst Center for Fundamental Interactions and Department of Physics, Amherst, MA 01003-9337 USA} \affiliation{Lawrence Berkeley National Laboratory, 1 Cyclotron Rd., Berkeley, CA 94720, USA} 
\author{D.P.~Hogan} \affiliation{University of California Berkeley, Department of Physics, Berkeley, CA 94720, USA}  
\author{M.~Horn} \affiliation{South Dakota Science and Technology Authority, Sanford Underground Research Facility, Lead, SD 57754, USA} \affiliation{University of California Berkeley, Department of Physics, Berkeley, CA 94720, USA} 
\author{D.Q.~Huang} 
\email[Corresponding Author:]{dongqing\_huang@alumni.brown.edu}
\affiliation{Brown University, Department of Physics, 182 Hope St., Providence, RI 02912, USA}%\footnote{Now at: University of Michigan, Randall Laboratory of Physics, 450 Church Street, Ann Arbor, MI 48109-1040, USA}
\author{C.M.~Ignarra} \affiliation{SLAC National Accelerator Laboratory, 2575 Sand Hill Road, Menlo Park, CA 94205, USA} \affiliation{Kavli Institute for Particle Astrophysics and Cosmology, Stanford University, 452 Lomita Mall, Stanford, CA 94309, USA} 
\author{R.G.~Jacobsen} \affiliation{University of California Berkeley, Department of Physics, Berkeley, CA 94720, USA}  
\author{O.~Jahangir} \affiliation{Department of Physics and Astronomy, University College London, Gower Street, London WC1E 6BT, United Kingdom}  
\author{W.~Ji} \affiliation{SLAC National Accelerator Laboratory, 2575 Sand Hill Road, Menlo Park, CA 94205, USA} \affiliation{Kavli Institute for Particle Astrophysics and Cosmology, Stanford University, 452 Lomita Mall, Stanford, CA 94309, USA} 
\author{K.~Kamdin} \affiliation{University of California Berkeley, Department of Physics, Berkeley, CA 94720, USA} \affiliation{Lawrence Berkeley National Laboratory, 1 Cyclotron Rd., Berkeley, CA 94720, USA} 
\author{K.~Kazkaz} \affiliation{Lawrence Livermore National Laboratory, 7000 East Ave., Livermore, CA 94551, USA}  
\author{D.~Khaitan} \affiliation{University of Rochester, Department of Physics and Astronomy, Rochester, NY 14627, USA}  
%\author{R.~Knoche} \affiliation{University of Maryland, Department of Physics, College Park, MD 20742, USA}  
\author{E.V.~Korolkova} \affiliation{University of Sheffield, Department of Physics and Astronomy, Sheffield, S3 7RH, United Kingdom}  
\author{S.~Kravitz} \affiliation{Lawrence Berkeley National Laboratory, 1 Cyclotron Rd., Berkeley, CA 94720, USA}  
\author{V.A.~Kudryavtsev} \affiliation{University of Sheffield, Department of Physics and Astronomy, Sheffield, S3 7RH, United Kingdom}  
%\author{N.A.~Larsen} \affiliation{Yale University, Department of Physics, 217 Prospect St., New Haven, CT 06511, USA}  
\author{E.~Leason} \affiliation{SUPA, School of Physics and Astronomy, University of Edinburgh, Edinburgh EH9 3FD, United Kingdom}  
%\author{C.~Lee} \affiliation{SLAC National Accelerator Laboratory, 2575 Sand Hill Road, Menlo Park, CA 94205, USA} \affiliation{Kavli Institute for Particle Astrophysics and Cosmology, Stanford University, 452 Lomita Mall, Stanford, CA 94309, USA} 
%\author{B.G.~Lenardo} \affiliation{University of California Davis, Department of Physics, One Shields Ave., Davis, CA 95616, USA} \affiliation{Lawrence Livermore National Laboratory, 7000 East Ave., Livermore, CA 94551, USA} 
\author{K.T.~Lesko} \affiliation{Lawrence Berkeley National Laboratory, 1 Cyclotron Rd., Berkeley, CA 94720, USA}  
%\author{C.~Levy} \affiliation{University at Albany, State University of New York, Department of Physics, 1400 Washington Ave., Albany, NY 12222, USA} \affiliation{Lawrence Berkeley National Laboratory, 1 Cyclotron Rd., Berkeley, CA 94720, USA} 
\author{J.~Liao} \affiliation{Brown University, Department of Physics, 182 Hope St., Providence, RI 02912, USA}  
\author{J.~Lin} \affiliation{University of California Berkeley, Department of Physics, Berkeley, CA 94720, USA}  
\author{A.~Lindote} \affiliation{LIP-Coimbra, Department of Physics, University of Coimbra, Rua Larga, 3004-516 Coimbra, Portugal}  
\author{M.I.~Lopes} \affiliation{LIP-Coimbra, Department of Physics, University of Coimbra, Rua Larga, 3004-516 Coimbra, Portugal}  
\author{A.~Manalaysay} \affiliation{Lawrence Berkeley National Laboratory, 1 Cyclotron Rd., Berkeley, CA 94720, USA} \affiliation{University of California Davis, Department of Physics, One Shields Ave., Davis, CA 95616, USA} 
\author{R.L.~Mannino} \affiliation{Texas A \& M University, Department of Physics, College Station, TX 77843, USA} \affiliation{University of Wisconsin-Madison, Department of Physics, 1150 University Ave., Madison, WI 53706, USA} 
\author{N.~Marangou} \affiliation{Imperial College London, High Energy Physics, Blackett Laboratory, London SW7 2BZ, United Kingdom}  
%\author{M.F.~Marzioni} \affiliation{SUPA, School of Physics and Astronomy, University of Edinburgh, Edinburgh EH9 3FD, United Kingdom}  
\author{D.N.~McKinsey} \affiliation{University of California Berkeley, Department of Physics, Berkeley, CA 94720, USA} \affiliation{Lawrence Berkeley National Laboratory, 1 Cyclotron Rd., Berkeley, CA 94720, USA} 
\author{D.-M.~Mei} \affiliation{University of South Dakota, Department of Physics, 414E Clark St., Vermillion, SD 57069, USA}  
%\author{J.~Mock} \affiliation{University at Albany, State University of New York, Department of Physics, 1400 Washington Ave., Albany, NY 12222, USA}  
%\author{M.~Moongweluwan} \affiliation{University of Rochester, Department of Physics and Astronomy, Rochester, NY 14627, USA}  
\author{J.A.~Morad} \affiliation{University of California Davis, Department of Physics, One Shields Ave., Davis, CA 95616, USA}  
\author{A.St.J.~Murphy} \affiliation{SUPA, School of Physics and Astronomy, University of Edinburgh, Edinburgh EH9 3FD, United Kingdom}  
\author{A.~Naylor} \affiliation{University of Sheffield, Department of Physics and Astronomy, Sheffield, S3 7RH, United Kingdom}  
\author{C.~Nehrkorn} \affiliation{University of California Santa Barbara, Department of Physics, Santa Barbara, CA 93106, USA}  
\author{H.N.~Nelson} \affiliation{University of California Santa Barbara, Department of Physics, Santa Barbara, CA 93106, USA}  
\author{F.~Neves} \affiliation{LIP-Coimbra, Department of Physics, University of Coimbra, Rua Larga, 3004-516 Coimbra, Portugal}  
\author{A.~Nilima} \affiliation{SUPA, School of Physics and Astronomy, University of Edinburgh, Edinburgh EH9 3FD, United Kingdom}  
%\author{K.~O'Sullivan} \affiliation{University of California Berkeley, Department of Physics, Berkeley, CA 94720, USA} \affiliation{Lawrence Berkeley National Laboratory, 1 Cyclotron Rd., Berkeley, CA 94720, USA} 
\author{K.C.~Oliver-Mallory} \affiliation{Imperial College London, High Energy Physics, Blackett Laboratory, London SW7 2BZ, United Kingdom} \affiliation{University of California Berkeley, Department of Physics, Berkeley, CA 94720, USA} \affiliation{Lawrence Berkeley National Laboratory, 1 Cyclotron Rd., Berkeley, CA 94720, USA}
\author{K.J.~Palladino} \affiliation{University of Wisconsin-Madison, Department of Physics, 1150 University Ave., Madison, WI 53706, USA}  
%\author{E.K.~Pease} \affiliation{University of California Berkeley, Department of Physics, Berkeley, CA 94720, USA} \affiliation{Lawrence Berkeley National Laboratory, 1 Cyclotron Rd., Berkeley, CA 94720, USA} 
%\author{L.~Reichhart} \affiliation{Department of Physics and Astronomy, University College London, Gower Street, London WC1E 6BT, United Kingdom}  
\author{C.~Rhyne} \affiliation{Brown University, Department of Physics, 182 Hope St., Providence, RI 02912, USA}  
\author{Q.~Riffard} \affiliation{University of California Berkeley, Department of Physics, Berkeley, CA 94720, USA} \affiliation{Lawrence Berkeley National Laboratory, 1 Cyclotron Rd., Berkeley, CA 94720, USA} 
\author{G.R.C.~Rischbieter} \affiliation{University at Albany, State University of New York, Department of Physics, 1400 Washington Ave., Albany, NY 12222, USA}  
\author{P.~Rossiter} \affiliation{University of Sheffield, Department of Physics and Astronomy, Sheffield, S3 7RH, United Kingdom}  
\author{S.~Shaw} \affiliation{University of California Santa Barbara, Department of Physics, Santa Barbara, CA 93106, USA} \affiliation{Department of Physics and Astronomy, University College London, Gower Street, London WC1E 6BT, United Kingdom} 
\author{T.A.~Shutt} \affiliation{SLAC National Accelerator Laboratory, 2575 Sand Hill Road, Menlo Park, CA 94205, USA} \affiliation{Kavli Institute for Particle Astrophysics and Cosmology, Stanford University, 452 Lomita Mall, Stanford, CA 94309, USA} 
\author{C.~Silva} \affiliation{LIP-Coimbra, Department of Physics, University of Coimbra, Rua Larga, 3004-516 Coimbra, Portugal}  
\author{M.~Solmaz} \affiliation{University of California Santa Barbara, Department of Physics, Santa Barbara, CA 93106, USA}  
\author{V.N.~Solovov} \affiliation{LIP-Coimbra, Department of Physics, University of Coimbra, Rua Larga, 3004-516 Coimbra, Portugal}  
\author{P.~Sorensen} \affiliation{Lawrence Berkeley National Laboratory, 1 Cyclotron Rd., Berkeley, CA 94720, USA}  
%\author{S.~Stephenson} \affiliation{University of California Davis, Department of Physics, One Shields Ave., Davis, CA 95616, USA}  
\author{T.J.~Sumner} \affiliation{Imperial College London, High Energy Physics, Blackett Laboratory, London SW7 2BZ, United Kingdom}  
\author{N.~Swanson} \affiliation{Brown University, Department of Physics, 182 Hope St., Providence, RI 02912, USA}  
\author{M.~Szydagis} \affiliation{University at Albany, State University of New York, Department of Physics, 1400 Washington Ave., Albany, NY 12222, USA}  
\author{D.J.~Taylor} \affiliation{South Dakota Science and Technology Authority, Sanford Underground Research Facility, Lead, SD 57754, USA}  
\author{R.~Taylor} \affiliation{Imperial College London, High Energy Physics, Blackett Laboratory, London SW7 2BZ, United Kingdom}  
\author{W.C.~Taylor} \affiliation{Brown University, Department of Physics, 182 Hope St., Providence, RI 02912, USA}  
\author{B.P.~Tennyson} \affiliation{Yale University, Department of Physics, 217 Prospect St., New Haven, CT 06511, USA}  
\author{P.A.~Terman} \affiliation{Texas A \& M University, Department of Physics, College Station, TX 77843, USA}  
\author{D.R.~Tiedt} \affiliation{University of Maryland, Department of Physics, College Park, MD 20742, USA}  
\author{W.H.~To} \affiliation{California State University Stanislaus, Department of Physics, 1 University Circle, Turlock, CA 95382, USA}  
%\author{M.~Tripathi} \affiliation{University of California Davis, Department of Physics, One Shields Ave., Davis, CA 95616, USA}  
\author{L.~Tvrznikova} \affiliation{University of California Berkeley, Department of Physics, Berkeley, CA 94720, USA} \affiliation{Lawrence Berkeley National Laboratory, 1 Cyclotron Rd., Berkeley, CA 94720, USA} \affiliation{Yale University, Department of Physics, 217 Prospect St., New Haven, CT 06511, USA}
\author{U.~Utku} \affiliation{Department of Physics and Astronomy, University College London, Gower Street, London WC1E 6BT, United Kingdom}  
%\author{S.~Uvarov} \affiliation{University of California Davis, Department of Physics, One Shields Ave., Davis, CA 95616, USA}  
\author{A.~Vacheret} \affiliation{Imperial College London, High Energy Physics, Blackett Laboratory, London SW7 2BZ, United Kingdom}  
\author{A.~Vaitkus} \affiliation{Brown University, Department of Physics, 182 Hope St., Providence, RI 02912, USA}  
\author{V.~Velan} \affiliation{University of California Berkeley, Department of Physics, Berkeley, CA 94720, USA}  
%\author{J.R.~Verbus} \affiliation{Brown University, Department of Physics, 182 Hope St., Providence, RI 02912, USA}  
\author{R.C.~Webb} \affiliation{Texas A \& M University, Department of Physics, College Station, TX 77843, USA}  
\author{J.T.~White} \affiliation{Texas A \& M University, Department of Physics, College Station, TX 77843, USA}  
\author{T.J.~Whitis} \affiliation{SLAC National Accelerator Laboratory, 2575 Sand Hill Road, Menlo Park, CA 94205, USA} \affiliation{Kavli Institute for Particle Astrophysics and Cosmology, Stanford University, 452 Lomita Mall, Stanford, CA 94309, USA} 
\author{M.S.~Witherell} \affiliation{Lawrence Berkeley National Laboratory, 1 Cyclotron Rd., Berkeley, CA 94720, USA}  
\author{F.L.H.~Wolfs} \affiliation{University of Rochester, Department of Physics and Astronomy, Rochester, NY 14627, USA}  
\author{D.~Woodward} \affiliation{Pennsylvania State University, Department of Physics, 104 Davey Lab, University Park, PA  16802-6300, USA}  
\author{X.~Xiang} \affiliation{Brown University, Department of Physics, 182 Hope St., Providence, RI 02912, USA}  
\author{J.~Xu} \affiliation{Lawrence Livermore National Laboratory, 7000 East Ave., Livermore, CA 94551, USA}  
%\author{K.~Yazdani} \affiliation{Imperial College London, High Energy Physics, Blackett Laboratory, London SW7 2BZ, United Kingdom}  
%\author{S.K.~Young} \affiliation{University at Albany, State University of New York, Department of Physics, 1400 Washington Ave., Albany, NY 12222, USA}  
\author{C.~Zhang} \affiliation{University of South Dakota, Department of Physics, 414E Clark St., Vermillion, SD 57069, USA}

\begin{abstract}
Dual-phase xenon time projection chamber (TPC) detectors offer heightened sensitivities for dark matter detection across a spectrum of particle masses. To broaden their capability to low-mass dark matter interactions, we investigated the light and charge responses of liquid xenon (LXe) to sub-keV nuclear recoils. Using neutron events from a pulsed Adelphi Deuterium-Deuterium neutron generator, an $\textit{in situ}$ calibration was conducted on the LUX detector. We demonstrate direct measurements of light and charge yields down to 0.45~keV and 0.27~keV, respectively, both approaching single quanta production, the physical limit of LXe detectors. These results hold significant implications for the future of dual-phase xenon TPCs in detecting low-mass dark matter via nuclear recoils.
\end{abstract}

\maketitle

\setcounter{equation}{0} \setcounter{footnote}{0} % what this does? RG - This takes of the numbering for footnotes, if you use them

%%%%%%%%%%%%%%%%%%%%%%%%%%%%%%%%%%%%%%%%%%%%%%%%%%%
\textit{Introduction}.\textemdash 
Dual-phase xenon time projection chambers (TPCs), a leading technology for dark matter detection\cite{Akerib:2016vxi, Aprile:2018dbl, PandaX-4T:2021bab, LZ:2022lsv}, measure nuclear recoils (NR) from weakly interacting massive particles (WIMPs) through both scintillation light (S1) and ionization charge (S2) in liquid xenon (LXe). Detecting low-mass dark matter remains challenging due to limited calibrations of low-energy NR responses. This study presents the first simultaneous measurements of light (L$_y$) and charge (Q$_y$) yields for NR in LXe, characterizing the average quanta per keV down to the sub-keV region using the Large Underground Xenon (LUX) detector. These yields were obtained indirectly by comparing data with simulation of the NR spectrum.

\textit{Data Collection and Analysis}.\textemdash In 2016, we enhanced the NR calibration of the LUX detector~\cite{LUXexperiment} \textit{in situ}, using neutron events from a pulsed Adelphi\footnote{Adelphi Technology Inc., 2003 E. Bayshore Road, Redwood City, CA 94063} Deuterium-Deuterium (D-D) neutron generator.\footnote{For the first LUX D-D neutron calibration (LUX DD2013) details, see~\cite{LUXDD, james_thesis, dd_brown}.}
LUX has a 250 kg active mass and 122 2-inch PMTs in top and bottom arrays, shielded by a 7.6 m × 6.1 m cylindrical water tank.
Incident particles generate immediate S1 scintillation photons, detected by PMTs with a gain (\(g_1\)) of \(0.096\pm0.003\) photon detected (phd)/photon~\cite{LUXRun3Reanalysisprl, DoublePhe}. Concurrently, the ionization charge drifts upwards in LXe and, upon transitioning to the gas phase, produces the S2 signal with an ionization gain (\(g_2\)) of \(18.5\pm0.9\) phd/electron. Each electron induces, on average, \(25.72\pm0.04\) phd with a width of \(5.47\pm0.03\) phd across PMTs~\cite{huang}. %The S1-S2 interval reveals the interaction depth (\(z\)) in LXe, and the S2 distribution on the top PMTs gives the \(xy\) event position. 
For LUX details, consult~\cite{LUXfirstResult, LUXRun3Reanalysisprl, run4fullexpo, spinDep, LUX:2018akb, LUX:2019gwa, LUX:2017glr, lux_xe127, LUXTritium, LUXDD, james_thesis}.

\begin{figure}
    \centering
    \includegraphics[width=0.45\textwidth]{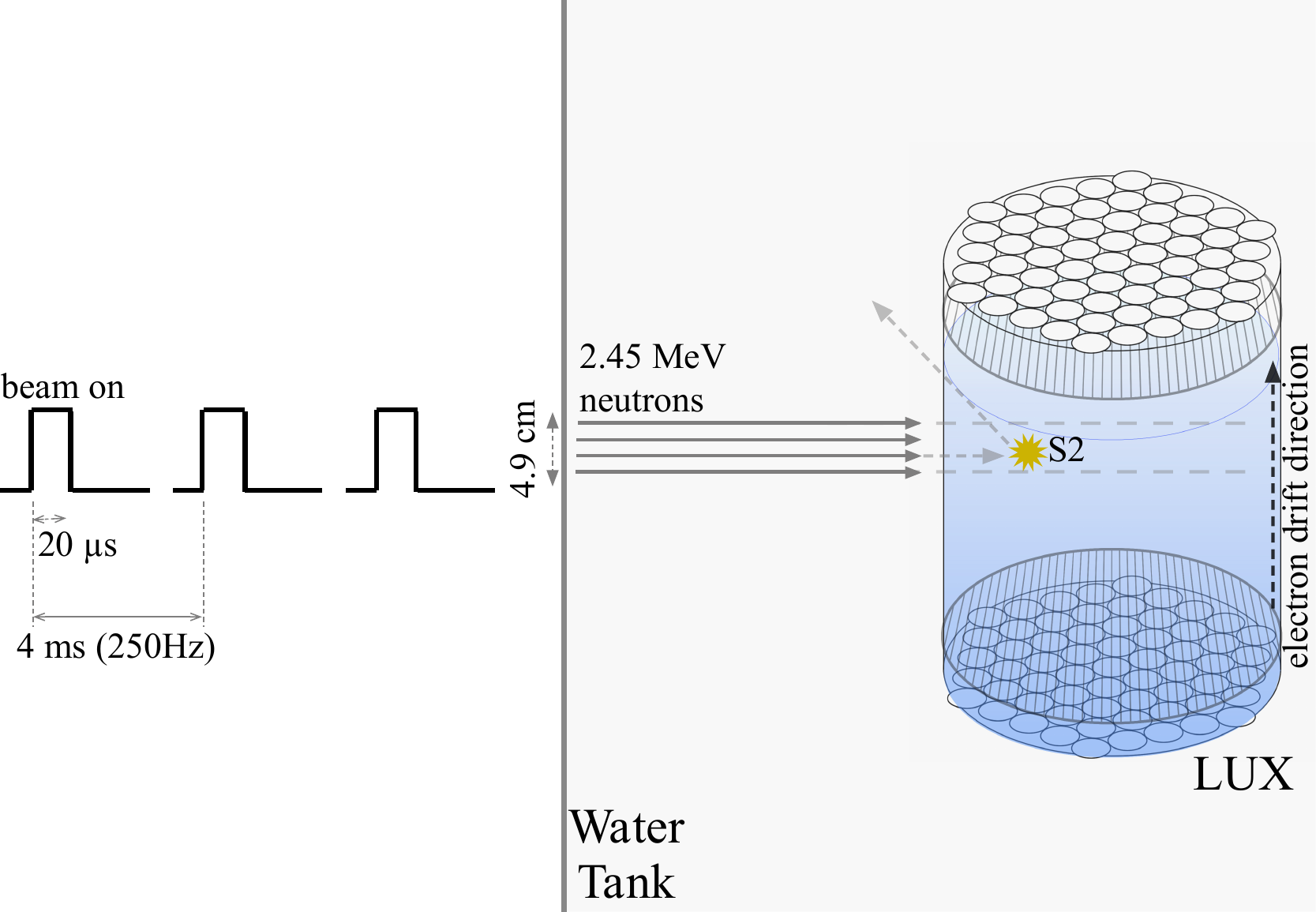}
    \caption{
    Diagram (not to scale) of the LUX's short-pulsed D-D neutron calibration.}
    \label{fig:schematic_setup}
\end{figure}

A schematic of the experimental setup is depicted in Fig.~\ref{fig:schematic_setup}. We directed a collimated neutron beam (2.45~MeV) through a conduit of 377~cm length and 4.9~cm diameter.
The conduit center is 10 cm below the LXe surface, within a 50~cm deep active volume.
The D-D generator operated at a 250~Hz frequency and a 20~$\upmu$s pulse width, producing an instantaneous flux of $2.8\times10^8$~neutrons/s. %Below 250~Hz, the generator becomes unstable. 
At this flux rate, on average, about 0.06 neutrons reach the TPC with each pulse, resulting in a probability of approximately $3\%$ for multiple neutron interactions per pulse. In the pulsed mode, the D-D generator's trigger time provides an estimate of the neutron interaction time in the TPC, enabling us to study low-energy events that produce detectable ionization signals without accompanying scintillation signals.

For yield measurements, we selected D-D neutron events that exhibited a single scatter, characterized by one observed S2 exceeding 44~phd; signals below this threshold are notably affected by spurious background single electrons (SE). This criterion may encompass neutron multi-scatters, where one S2 exceeds the threshold and others do not, we address this potential systematics in our signal modeling. Targeting low-energy neutron-induced xenon recoils, we permitted events with zero or one preceding S1 pulse to the S2. In this context, an S1, unlike those in other LUX analyses, is defined as a scintillation signal without the typical two-fold coincidence, with its magnitude quantified by the discrete photon counts on the PMTs, termed `spikes'~\cite{LUXRun3Reanalysisprl}. 
S2 signals must occur within 65 to 125 µs after a D-D trigger, align with the neutron conduit depth~(7.5-12.5 cm), and be located within the neutron beam's $xy$ projection, defined as a 7 cm diameter cylinder, to capture the majority of signal events while eliminating spurious coincidences. For events with S1, a time cut of $>2.5~\upmu$s between S1 and S2 further refines our selection, eliminating events where S1 pulses are misconstrued from the leading edge of an S2. To maximize event inclusion, we abstain from a radial fiducialization cut. Notably, S2 signal charge loss near the TPC wall is deemed negligible (0.13\% of events)~\cite{huang}, attributed to charge accumulation on the wall, guiding signals inward during vertical transit~\cite{run4fullexpo}.

\begin{figure}
    \centering
    \includegraphics[width=0.45\textwidth]{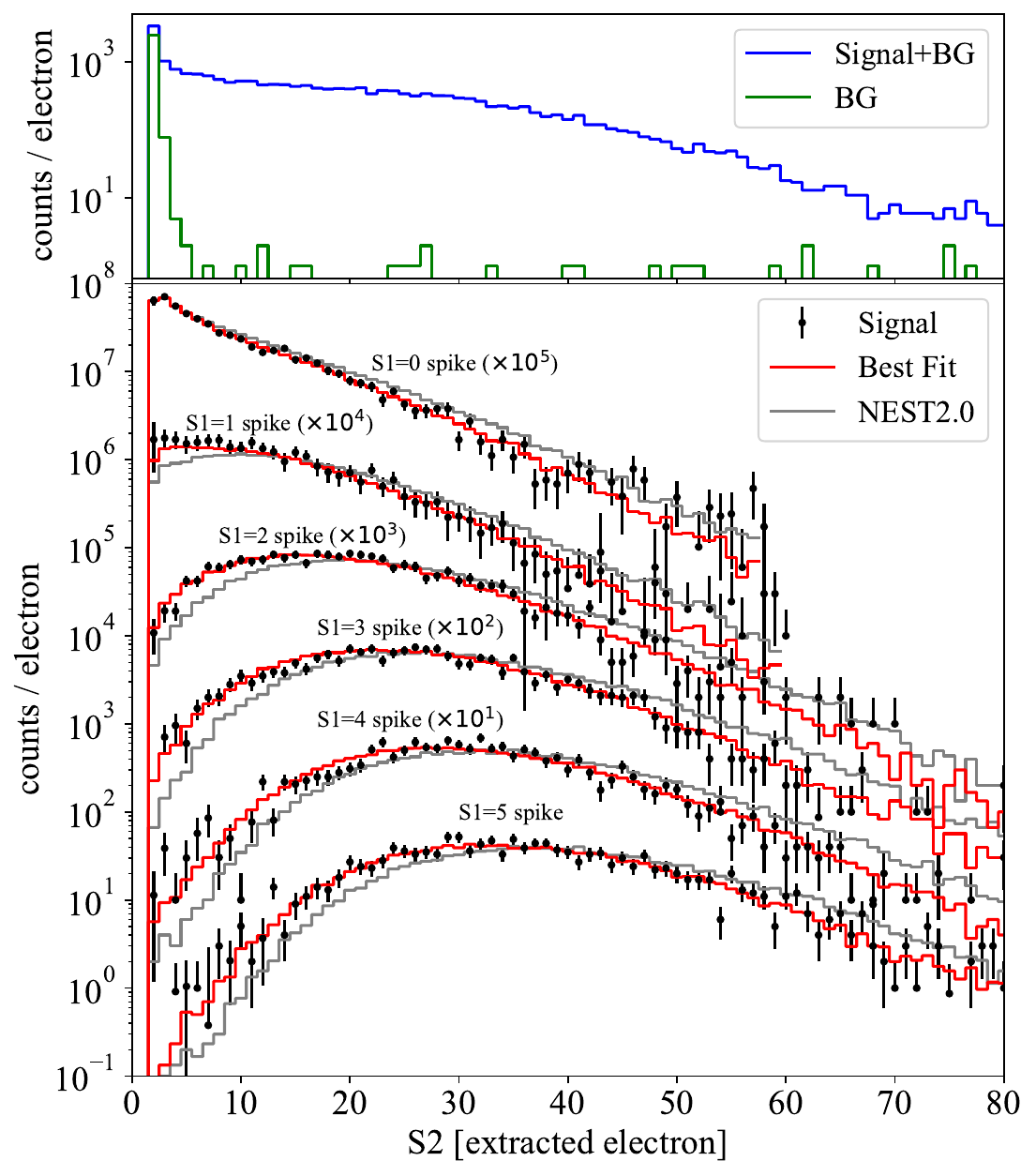}
    \caption{Top panel: The blue histogram shows the combined S2 spectrum of data for S1 values ranging from 0 to 5 spikes. The green histogram represents the measured background from events where the S2 precedes the D-D trigger pulse within the same dataset. Bottom panel: Background-subtracted S2 spectra corresponding to S1 spikes of 0 to 5 are shown as black points. The red histograms represent the best-fit results, while the gray ones are produced from the original NEST2.0 yield models. For display purposes, both measured and modeled S2 spectra for S1 = $i$ spike have been scaled by a factor of $10^{5-i}$, where $i = 0, 1, 2, 3, 4, 5$. Histogram bins and error bars for black data points extending into the negative domain have been suppressed.
}
    \label{fig:chap8_best_fit_vs_nest2p0_updated_prl}
\end{figure}

The primary background in this study stems from electron-train (e-train) events, ubiquitous in xenon TPCs. Defined as sequences of single or clustered few-electron emissions trailing large S2 pulses with roughly 10~ms time constants~\cite{jingke_electron}, these e-trains may be mistakenly identified as S2 signals from low-energy neutron interactions, complicating the calibration process. %Leveraging the D-D trigger's temporal precision drastically curtails this e-train interference. 
Utilizing the temporal precision of the D-D trigger to require coincidence with the TPC signals effectively eliminates prevalent e-train background interference. Two additional quiet-time cuts further diminish e-train contamination: the first mandates a 4 ms hiatus between LUX-triggered events and the candidate signal, and the second asserts that no SE emissions precede the observed S2 within the event. Both cuts, optimized for signal-to-noise ratio, reduce e-train events by factors of three and two, respectively, with $80\%$ signal acceptance. The `no-SE-ahead-S2' cut, however, might inadvertently exclude genuine events that exhibit SE from low-energy neutron interactions, potentially compromising signal uptake.
%Nonetheless, the no-SE-ahead-S2 condition could inadvertently snub genuine low-energy neutron scatter events, compromising signal uptake. 
This bias will be addressed in subsequent modeling. Moreover, the D-D trigger facilitates \textit{in situ} evaluations of the lingering e-train event rate by probing TPC events where the S2 appears before the D-D trigger pulse, as shown in Fig.~\ref{fig:chap8_best_fit_vs_nest2p0_updated_prl} (top panel). The background rate for events featuring a D-D neutron S1 but lacking a corresponding S2, which instead coincide with e-train S2s, is quantified using the NEST2.0 model~\cite{szydagis_m_2018_4062516}. This background predominantly affects the lowest-energy bins and is continuously updated along with the NEST2.0 yield models when fitting the signal model to the data, which will be discussed later.

Random small S1s, primarily photoelectron (PHE) pulses from PMT dark counts or subsequent to high energy depositions in the TPC~\cite{jingke_electron}, pose major challenges in accurately identifying signal events with 0-spike and 1-spike S1s. The average background PHE rate is $1.8\pm0.1$ within 1-ms event windows, complicating data interpretation. For instance, a coincidental PHE pulse aligning with a 0-spike event could cause the event to be misinterpreted as a 1-spike event. Additionally, within genuine 1-spike events, extraneous background PHEs can give the appearance of multiple 1-spike S1 pulses preceding the primary S2 signal. %However, a distinct temporal pattern enables the statistical differentiation of authentic 1-spike S1 signals from background PHEs. 
Unlike the uniform temporal distribution of background PHEs, genuine 1-spike S1 signals from D-D neutron interactions are concentrated within a narrowly defined D-D S1 window prior to S2 emergence. This temporal distinction, coupled with D-D trigger timing, enables us to statistically distinguish between signal events and accidentals, and to discern between 0-spike and 1-spike S1 signal events by analyzing their collective temporal distributions. For details, see Sec.~6.3.4 in~\cite{huang}.

Upon completing the event selection and background analysis, we determined the absolute rates for D-D neutron single elastic scatter events with S1 spanning $0,1,2,3,4,5$ spikes. Figure~\ref{fig:chap8_best_fit_vs_nest2p0_updated_prl} (bottom panel) showcases the S2 spectra corresponding to each S1 (represented by black data points). S2 pulse areas, for events with S1~$\geq2$ spikes, are corrected for position-dependent detection efficiency using a $^{83m}$Kr calibration. This correction is not implemented to S2s associated with 0 and 1-spike S1s due to the lack of accurate $z$-position information, resulting from either the absence of S1 signals or confusion caused by background PHEs. For consistency, the same treatment of the S2 pulse area is applied in the signal modeling.

\textit{Signal Modeling}.\textemdash To model the differential NR spectra arising from single elastic scatter interactions of neutrons at low energies, we conducted a GEANT4-based simulation, which fully incorporates the LUX geometry, including the LUX water tank and the D-D neutron conduit~(LUXSim~\cite{luxsim_paper}). In this simulation, we exclusively select neutron events involving either elastic scattering or neutron capture~\footnote{Radiative neutron capture by xenon isotopes results in finite NR energy deposition (up to 0.3 keV) in LXe~\cite{huang, Amarasinghe:2022jgk}.}, while vetoing any events with gamma-ray energy depositions.
For each simulated neutron event, we record the four highest-energy deposition vertices, capturing crucial details such as neutron and deposited energies, along with their $(x, y, z)$ positions. This dataset enabled us to model systematic effects related to event selection criteria, including the S2 threshold and the `no-SE-ahead-S2' cut, as well as the merging of adjacent S2 pulses in the vertical direction.

During this calibration, the LUX active volume exhibited a notably non-uniform electric field~\cite{run4fullexpo}. To account for this field variation, a dedicated field model~\cite{lux_field} was specifically developed for the calibration period. This field model is utilized to calculate the electric field strength at each recorded vertex. Since our analysis is carried out in the observed space, we employ the same field model to map each simulated vertex from real space into observed space. The weighted average of the electric field for selected neutron events is $400\pm80$~V/cm.

A LUX-adapted NEST2.0 program is employed to simulate the production and detection of S1 and S2 signals for each recorded vertex, utilizing information derived from deposited energy, electric field strength, and position. At the core of NEST2.0 lie the empirically-derived L$_y$ and Q$_y$ models. The recoil interaction initially generates N$_\textrm{ex}$ excitons (Xe$^*$) and N$_\textrm{i}$ electron-ion (e$^-$Xe$^+$) pairs at the interaction site. These excitations subsequently de-excite or recombine, resulting in the production of S1 and S2 signals. The fluctuations in N$_\textrm{ex}$ and N$_\textrm{i}$ are independently modeled using Gaussian statistics, with widths ($\sigma$) determined by $\sqrt{F\overline{N}_\textrm{ex}}$ and $\sqrt{F\overline{N}_\textrm{i}}$, respectively, within NEST2.0. Here, $F$ represents a Fano-like factor. While the value of $F$ is consistent with 1 based on DD2013~\cite{LUXDD, james_thesis} and the XENON10 AmBe calibration~\cite{Sorensen2009339}, it carries a significant uncertainty due to the absence of mono-energetic lines in NR calibrations. The treatment of $F$ is discussed in the next section.

LUX detector parameters were measured \textit{in situ} for the calibration period~\cite{huang, LUXprd}. 
A single photon would lead to 1.17~PHE and the single PHE resolution is $1.00\pm0.46$.
Other measured parameters include the SE mean pulse area and width, and \(g_1\) and \(g_2\) as presented earlier. These parameters were incorporated into NEST2.0 for simulating signal detection processes. Following signal detection modeling, we combined any two S2 signals in a simulated neutron event with a $z$ separation of $<2~\upmu$s~(D-D neutron S2 1--99$\%$ width) in drift time. To determine the S1 to D-D trigger time for each vertex, we sampled it from a time distribution directly measured from data and added it to the drift time of each vertex, obtaining the S2 to D-D trigger time. The S2 trigger efficiency of the data acquisition system, measured from a separate D-D calibration dataset~\cite{LUXTriggerEfficiency}, was applied to S2s in each simulated event for event triggering. We corrected the S2 pulse area of surviving events with S1~$\geq2$ spikes to match real data. Additionally, we evaluated S1 pulse finding and classification efficiencies as functions of size through visual assessment of 6000 events using calibration data. These measured efficiencies were applied to simulated events for consistency with real data. All simulated events underwent the same event selection criteria as real data. The resulting signal model is presented in Fig.~\ref{fig:chap8_best_fit_vs_nest2p0_updated_prl} (gray histograms). A noticeable discrepancy between the calibration data and the original signal model was observed, which may be attributed to the limited constraints on yields at very low energies from~\cite{LUXDD, james_thesis}, the primary basis for NEST2.0 yield models.

\begin{figure}
    \centering
    \includegraphics[width=0.45\textwidth]{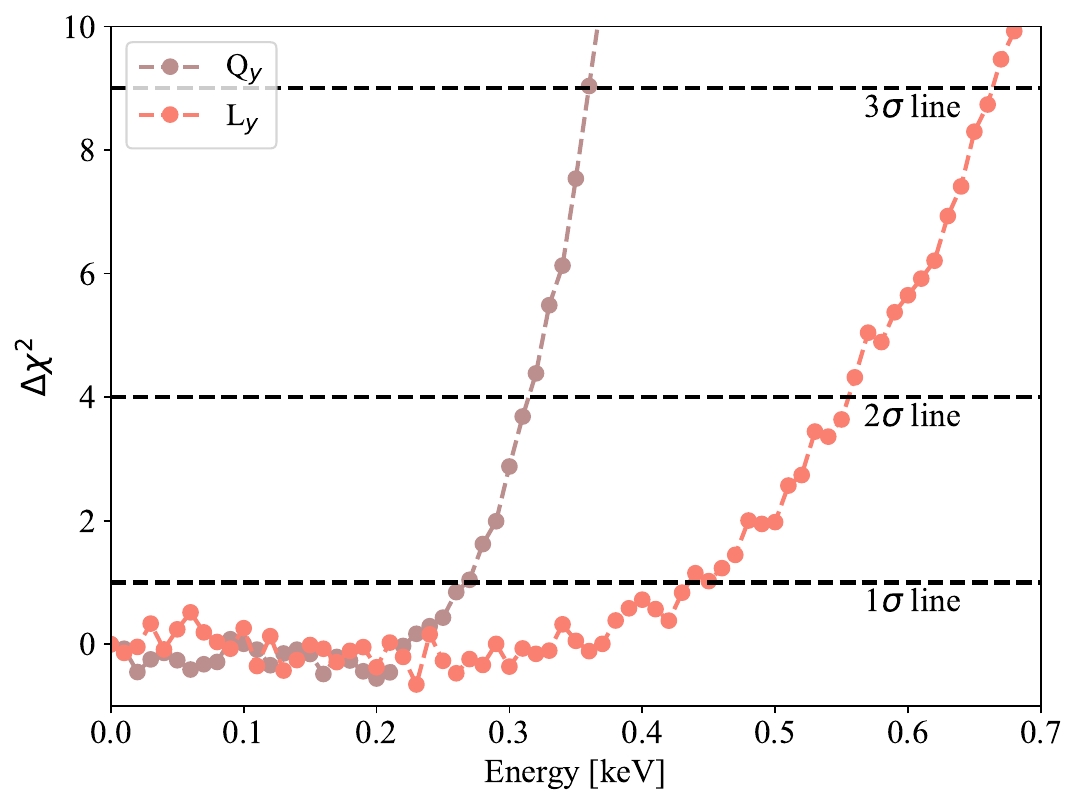}
    \caption{Evaluated $\Delta\chi^2$ values as a function of energy. The lowest energies to which these calibration data are sensitive to for L$_y$ and Q$_y$ are $0.45\pm0.03$~keV and $0.27\pm0.04$~keV at 1-$\sigma$ sensitivity level, respectively. At 2-$\sigma$ level, L$_y$ $0.56\pm0.02$~keV, Q$_y$ $0.31\pm0.03$~keV; and at 3-$\sigma$ level, L$_y$ $0.66\pm0.02$~keV, Q$_y$ $0.35\pm0.03$~keV.}
    \label{fig:threshold}
\end{figure}

\textit{Yield Measurements}.\textemdash Following detailed signal modeling, we adjusted the light yield (L$_y$) and charge yield (Q$_y$) models within NEST2.0—both shapes and amplitudes down to 0 keV—simultaneously yet independently, to achieve the best fit to the calibration data~(Fig.~\ref{fig:chap8_best_fit_vs_nest2p0_updated_prl}) using the least squares method:
\[
\chi^2 = \sum_{i=1}^{N} \frac{(O_i - E_i)^2}{E_i},
\]
where $O_i$ and $E_i$ are the observed and simulated counts per bin, and $N$ is the total number of valid bins across the six spectra.
The shape adjustments primarily focus on the low-energy end ($<3$~keV), as the high-energy end is well constrained by DD2013~\cite{LUXDD}. The L$_y$ primarily affects the relative counts and shapes of the six S2 spectra, whereas Q$_y$ influences their shapes. The parametrization of the yield models is illustrated in Figure 8.7 and 8.8 of~\cite{huang}. In the simultaneous fitting of these six S2 spectra, we employ a single overall event rate normalization factor to enforce a robust constraint. This factor is intentionally left free to ensure conservativeness.
The Fano-like factor $F$ is used to adjust the S2 spectrum widths in the signal model. It is treated as a free parameter in the fitting due to its unknown uncertainty. This conservative approach also captures other secondary factors contributing to the signal distribution widths. The best fit is achieved with $\chi^2 = 246.0$ for 262 degrees of freedom~($N_{\text{dof}}$) and 8 parameters. The $\chi^2/N_{\text{dof}}$ values for the S1 = 0, 1, 2, 3, 4, and 5 spike S2 spectra are 41.9/36, 17.1/30, 47.1/38, 53.2/43, 39.3/31, and 47.4/44, respectively.
%The best fit is achieved at $\chi^2 = 211.4$ with 243 degrees of freedom and 8 parameters. 
%The uncertainties on both L$_y$ and Q$_y$ are conservatively determined by marginalizing over all other fitting parameters, capturing systematics due to degeneracy between L$_y$ and Q$_y$ in the fitting. 
The fitting results indicate a positive correlation between L$_y$ and Q$_y$. Uncertainties on both L$_y$ and Q$_y$ are conservatively determined by marginalizing over all other fitting parameters, capturing the effects of their mutual correlation within the fit.
As $g_1$ and $g_2$ are in direct degeneracy with both L$_y$ and Q$_y$ for the observed S1 and S2 distributions, we assess the contributions of their non-negligible uncertainties to the yield measurements by repeating the fitting using $g_1$ and $g_2$ values at their 1-$\sigma$ uncertainty levels. For details, see Sec.~8.2 in~\cite{huang}. %For more on the fitting procedures, refer to Section 8.2.2 in~\cite{huang}.

\begin{figure}
    \centering
    \includegraphics[width=0.45\textwidth]{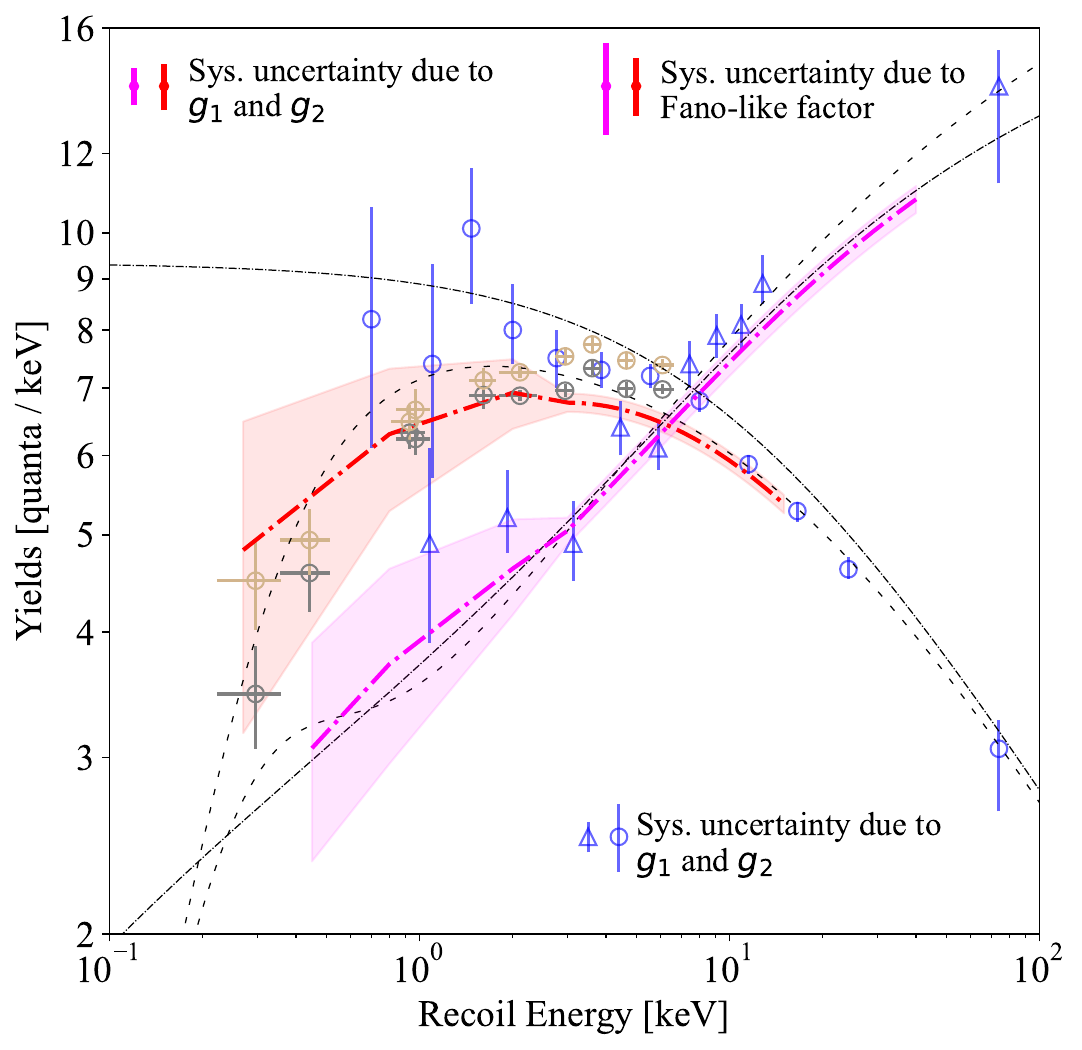}
    \caption{L$_y$ (magenta) and Q$_y$ (red) measurements for DD2016 at 400 V/cm are shown with statistical uncertainties (light bands). Q$_y$ results from XeNu 2019 at 220 V/cm (gray) and 550 V/cm (golden)~\cite{Lenardo:2019vkn}, and DD2013 at 180 V/cm (light blue triangles for L$_y$, circles for Q$_y$) are included. NEST2.0 and NEST2.3.11 models at 400 V/cm are shown as black dash-dotted and dashed lines, respectively. Code to extract and implement the DD2016 results is available at \url{https://gitlab.com/huangdq2017/implementation_of_lux_dd_yields}.}
    \label{fig:yields}
\end{figure}

To establish the lowest energies to which this calibration data is sensitive for L$_y$ and Q$_y$, we independently evaluate them based on the best-fit yield models. This involves cutting off the corresponding yield (assume zero yield) below certain energies from the best-fit yield model and calculating the $\Delta\chi^2$ values concerning the case of no energy cutoff in the signal model against the calibration data. The results, along with 1-$\sigma$, 2-$\sigma$, and 3-$\sigma$ sensitivity lines, are shown in Fig.~\ref{fig:threshold}. 
%The quoted uncertainties include systematics due to the Fano-like factor. 
We report the lowest energies that the data are sensitive to for both L$_y$ and Q$_y$ at the 1-$\sigma$ sensitivity level, yielding L$_y$ and Q$_y$ measurements of $0.45\pm0.03$~keV and $0.27\pm0.04$~keV, respectively, representing the lowest-energy NR calibrations in LXe to date. The final L$_y$ and Q$_y$ measurements of this work (DD2016) are presented in Fig.~\ref{fig:yields}.

With both L$_y$ and Q$_y$ measurements, we can constrain the Lindhard model~\cite{Sorensen2011, Lindhard1963}, which describes the quenching of electronic excitation from NR in LXe. The Lindhard factor $k$ is measured to be $0.146\pm0.013$, assuming a constant $W$ value~(energy required to produce a scintillation or ionization quantum) of 13.7~eV~\cite{Dahl:2009nta}. This value is consistent with the standard Lindhard model prediction of 0.166 within 1.5-$\sigma$. %The DD2013 measurement is $0.174\pm0.006$~\cite{LUXDD}, a higher value primarily due to weak constraints at the low-energy end. 
% (Fig.~\ref{fig:yields}).

\begin{figure}
    \centering
    \includegraphics[width=0.45\textwidth]{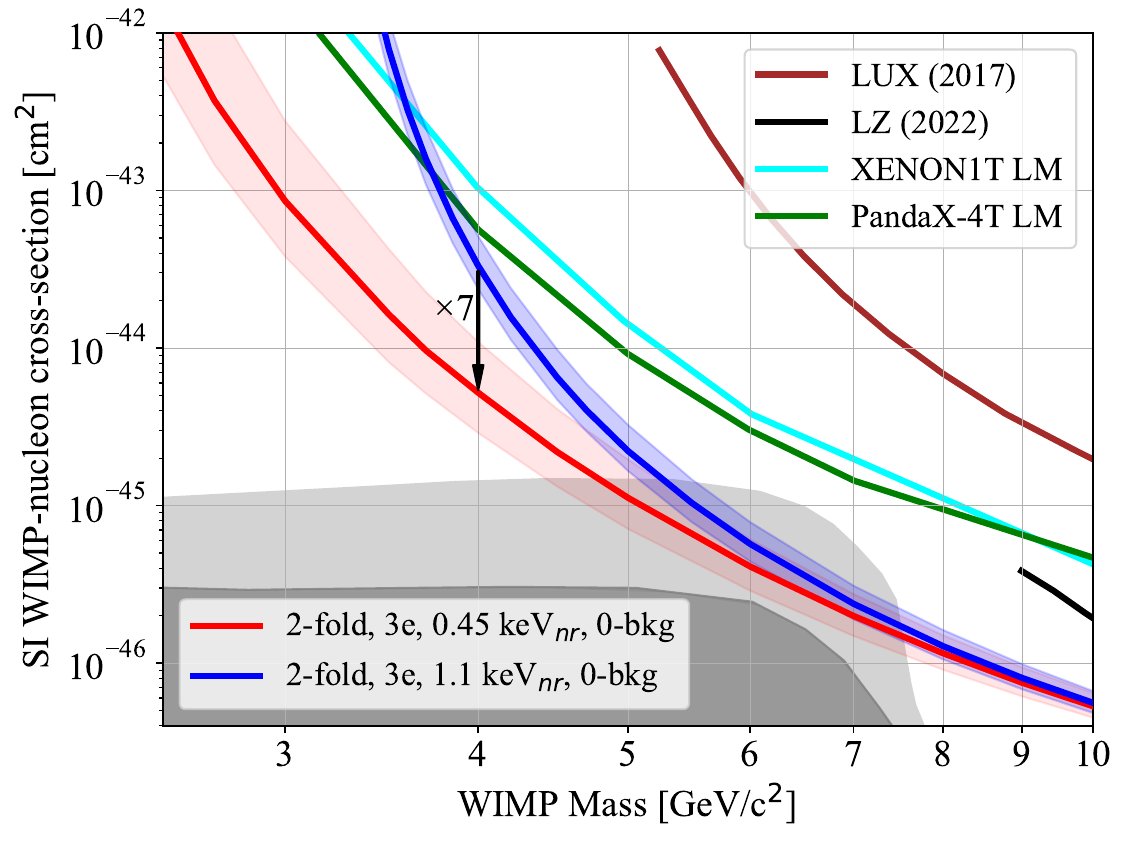}
    \caption{The projected $90\%$ sensitivities for a background-free LXe experiment with LZ-SR1~\cite{LZ:2022lsv} exposure are shown as red and blue curves for different energy thresholds, with uncertainty bands reflecting yield uncertainties. The searches use both scintillation and ionization channels with a 2-fold coincidence requirement and a three extracted-electron threshold. The curves demonstrate the low-mass WIMP search sensitivity improvement due to the lower NR energy threshold obtained in this work. For reference, limits from LUX (2017) (brown)\cite{run4fullexpo}, XENON1T (2021) low-mass (LM) WIMP search (cyan)\cite{XENON:2020gfr}, PandaX-4T (2023) LM WIMP search (green)\cite{PandaX:2022aac}, and LZ (2022) (black)\cite{LZ:2022lsv} are also shown. The gray regions represent the neutrino floors for the baseline scenario (light) and after reducing solar flux uncertainties by a factor of 10 (dark)~\cite{OHare:2021utq}.}
    \label{fig:limit_curve}
\end{figure}

%\section{Implication for low-mass WIMP search}\label{sec:Implication for low-mass WIMP search}
\textit{Impact on WIMP search and beyond}.\textemdash 
This work allows us to estimate the potential sensitivity of an optimized dual-phase xenon TPC to low-mass WIMP interactions within the standard halo model~\cite{Lewin:1995rx}. Figure~\ref{fig:limit_curve} demonstrates the gain in sensitivity for low-mass WIMP dark matter, benefiting from the improvements in the calibration of signal yields presented here. The limit curves are generated using NEST2.3.11~\cite{nest2_3_11}, with light and charge yield models matching this work. The search includes both S1 and S2 channels, with a two-fold PMT coincidence requirement and a three extracted-electron threshold. A background-free 0.9~tonne-year exposure, equivalent to the LZ-SR1 exposure~\cite{LZ:2022lsv}, is assumed. Zero WIMP acceptance is enforced for recoil energies below 0.45~keV and 1.1~keV, corresponding to the lowest yield measurements of this work and~\cite{LUXDD}, respectively. Sensitivity improvements greater than a factor of $\times7$ are achieved for WIMP masses below 4~GeV/c$^2$. A thorough investigation of detector accidental coincidence backgrounds, combined with leveraging the double-PHE effect~\cite{LUXdpe,LZdpe}, is necessary to achieve S1 and S2 thresholds at this level. As xenon TPCs improve in sensitivity for dark matter detection, lowering the energy threshold toward the Solar Boron-8 neutrino floor becomes increasingly critical~\cite{XENON:2020gfr, PandaX:2022aac, XENON:2024ijk, PandaX:2024muv}. This work also broadens the application of dual-phase TPCs to other rare event searches, such as coherent elastic neutrino-nucleus scattering from reactor antineutrinos~\cite{RELICS:2024opj} and probing a potential nonzero neutrino magnetic moment~\cite{Scholberg:2005qs}. Additionally, achieving near-single-quantum sensitivity in both light and charge yields marks a significant technological milestone, offering deeper insights into noble liquid microphysics.

%\section{Acknowledgements}\label{sec:Acknowledgements}
This work was partially supported by the U.S. Department of Energy (DOE) under Award No. DE-AC02-05CH11231, DE-AC05-06OR23100, DE-AC52-07NA27344, DE-FG01-91ER40618, DE-FG02-08ER41549, DE-FG02-11ER41738, DE-FG02-91ER40674, DE-FG02-91ER40688, DE-FG02-95ER40917, DE-NA0000979, DE-SC0006605, DE-SC0010010, DE-SC0015535, and DE-SC0019066; the U.S. National Science Foundation under Grants No. PHY-0750671, PHY-0801536, PHY-1003660, PHY-1004661, PHY-1102470, PHY-1312561, PHY-1347449, PHY-1505868, and PHY-1636738; the Research Corporation Grant No. RA0350; the Center for Ultra-low Background Experiments in the Dakotas (CUBED); and the South Dakota School of Mines and Technology (SDSMT).
Laborat\'{o}rio de Instrumenta\c{c}\~{a}o e F\'{i}sica Experimental de Part\'{i}culas (LIP)-Coimbra acknowledges funding from Funda\c{c}\~{a}o para a Ci\^{e}ncia e a Tecnologia (FCT) through the Project-Grant PTDC/FIS-NUC/1525/2014. Imperial College and Brown University thank the UK Royal Society for travel funds under the International Exchange Scheme (IE120804). The UK groups acknowledge institutional support from Imperial College London, University College London, the University of Sheffield, and Edinburgh University, and from the Science \& Technology Facilities Council for PhD studentships R504737 (EL), M126369B (NM), P006795 (AN), T93036D (RT) and N50449X (UU). This work was partially enabled by the University College London (UCL) Cosmoparticle Initiative. The University of Edinburgh is a charitable body, registered in Scotland, with Registration No. SC005336.
This research was conducted using computational resources and services at the Center for Computation and Visualization, Brown University, and also the Yale Science Research Software Core.
We gratefully acknowledge the logistical and technical support and the access to laboratory infrastructure provided to us by SURF and its personnel at Lead, South Dakota. SURF was developed by the South Dakota Science and Technology Authority, with an important philanthropic donation from T. Denny Sanford. SURF is a federally sponsored research facility under Award Number DE-SC0020216.

%\section{acknowledgements} \label{sec:acknowledgements}

%\bibliographystyle{apsrev4-1}

\bibliography{biblist}

\end{document}